\documentclass[onecolumn,authoryear]{els-mrw}

\usepackage{amsmath,amssymb,amsfonts,amsthm,makeidx,graphicx}
\usepackage{txfonts}
\usepackage{helvet}

\usepackage{graphicx}
\usepackage{graphics}
\usepackage{amssymb}
\usepackage{amsmath}
\usepackage{subfigure}
\usepackage{epsfig}
\usepackage{color}
\usepackage{array}
 
\usepackage{float}
\usepackage{soul}
\usepackage{wrapfig}
 \usepackage{stmaryrd}
 \usepackage{multicol}
\usepackage{multirow}
\usepackage{caption}

\newtheorem{Assumption 1}{Assumption}
\newtheorem{Definition 1}{Definition}
\usepackage{multirow}

\begin{document}

\chapter{Privacy in Multi-agent Systems}\label{chap1}

\author[1]{Yongqiang Wang}%


\address[1]{\orgname{Clemson University}, \orgdiv{Department of Electrical and Computer Engineering}, \orgaddress{Clemson, SC, 29634}}


\maketitle

%
%
%

\begin{abstract}[Abstract]
With the increasing awareness of privacy and the deployment of legislations in various multi-agent system application domains such as power systems and intelligent transportation, the privacy protection problem for multi-agent systems is gaining increased traction in recent years. This article discusses some of the representative advancements in the filed.
\end{abstract}

\section{Introduction}

All distributed   algorithms for multi-agent systems require the sharing of information across the agents. The  information sharing, although crucial to fulfill the coordination objective in multi-agent systems, also poses a threat for the privacy of participating agents  in applications involving sensitive data. For example, in
the rendezvous problem where a group of robots use  distributed
optimization to cooperatively find an optimal assembly point,
participating robots may want to keep their initial positions
private, which is particularly important in unfriendly
environments~\citep{zhang2019admm}.
In sensor network based localization, the positions of sensor agents should be kept private  in sensitive (hostile) environments as well~\citep{zhang2017distributed,zhang2019admm,huang2015differentially}.
In fact, without an effective privacy
mechanism in place, the results
in~\cite{zhang2019admm,huang2015differentially,burbano2019inferring}
show  that a participating agent's  
position  can be easily  inferred by an adversary or other
participating agents in distributed-optimization based rendezvous
and localization approaches. In multi-agent social networks, the opinions of individuals should also be kept private in many scenarios \citep{ye2019influence}.
Another example underscoring the importance
of privacy protection in multi-agent systems is distributed machine
learning where exchanged data may contain sensitive information such
as medical records or salary information~\citep{yan2012distributed}.
In fact, recent  results in~\cite{zhu2019deep} show  that without a
privacy mechanism in place, an adversary can use shared information to precisely recover the raw data used for training
(pixel-wise accurate for images and token-wise matching for texts).

Although plenty of privacy mechanisms have been developed in the computer science domain, including differential privacy \citep{dwork2014algorithmic}, cryptography, secure-multiparty computation, etc, those mechanisms are developed for {\it static} data. Therefore, when directly applied to multi-agent systems involving {\it dynamics}, those privacy mechanisms usually fall short due to excessive computation/communication overhead or loss of algorithmic accuracy. In the past few years, plenty of efforts have been devoted to privacy protection in multi-agent systems. This article discusses some of the typical results in the control  domain.  It is worth noting that due to the  vast amount of publications in this area in the past several years, our discussions  do  not pretend to be exhaustive and we apologize to anyone whose work is left out or not given the attention it deserves.

We consider two types of adversaries:

\emph{ An honest-but-curious
    adversary}  is an agent who
 follows all protocol steps correctly but is curious and collects
 received  data in an attempt to learn some information about other
 participating agents.

\emph{An eavesdropper} is an external attacker   who knows the
network topology, and is able to wiretap communication links and
access exchanged messages.

Generally speaking, an eavesdropper is more disruptive than an
honest-but-curious agent in terms of information breaches because it
can snoop messages exchanged on many channels whereas the latter can
only access the messages destined to it. However, an
honest-but-curious agent does have one piece of information that is
unknown to an external eavesdropper, i.e., the internal  
state   information of agent $i$ is available to the adversary if agent $i$ is an honest-but-curious
agent.  

We will consider three typical algorithms that underpin  most multi-agent applications, i.e., the static average consensus, the dynamic average consensus, and distributed optimization. We will use agents and nodes interchangeably. 

\section{Privacy protection for static average consensus}

\subsection{Problem formulation}

\paragraph{Static average consensus} Usually, the static average consensus  is also called average consensus. Following the
convention  in \cite{Olfati-Saber2007}, we represent a network of
$m$ nodes as  a graph ${G=(V,\,E,\,L)}$ with node set
${V}=\{v_1,\,v_2,\,\cdots, v_m\}$, edge set ${E}\subset {V}\times
{V}$, and the adjacency matrix $L=\big[L_{ij}[k]\big]$
denoting coupling weights which satisfy $L_{ij}[k]>0$ if
$(v_i,v_j)\in E$ and 0 otherwise. Here $k$ is time index, denoting
that $L_{ij}[k]$ could be time-varying. The set of neighbors of a
node $v_i$ is denoted as $\mathbb{N}_i = \left\{v_j \in {V}| (v_i,v_j)\in
{E}\right\}$ and its cardinality is denoted as $|\mathbb{N}_i|$.

%

 We represent the state variable of a node $i$ as $x_i[k]$. For the sake of expositional simplicity, we assume scalar states. But
 the results are easily extendable to the case where the state is a
 vector.
  To
achieve average consensus, namely convergence of all states $x_i[k]$
$(i=1,2,\cdots,m)$ to the average of initial values, i.e.,
$\frac{\sum_{i=1}^m x_i[0]}{m}$, the update rule    is   formulated
as \citep{olfati2007consensus}
\begin{equation}
\label{eq:dt} x_i[k+1] = x_i[k] + \varepsilon\sum_{v_j\in \mathbb{N}_i}
L_{ij}[k] (x_j[k]- x_i[k])
\end{equation}
where $\varepsilon$ resides in the range $(0, \frac{1}{\Delta}]$
with $\Delta$ defined as
\begin{equation}\label{eq:Delta}
\Delta\triangleq\max_{i=1,2,\cdots,m}|\mathbb{N}_i|
\end{equation}
\begin{wrapfigure}{R}{0.6\textwidth}
 \centering
 \includegraphics[width=.6\textwidth]{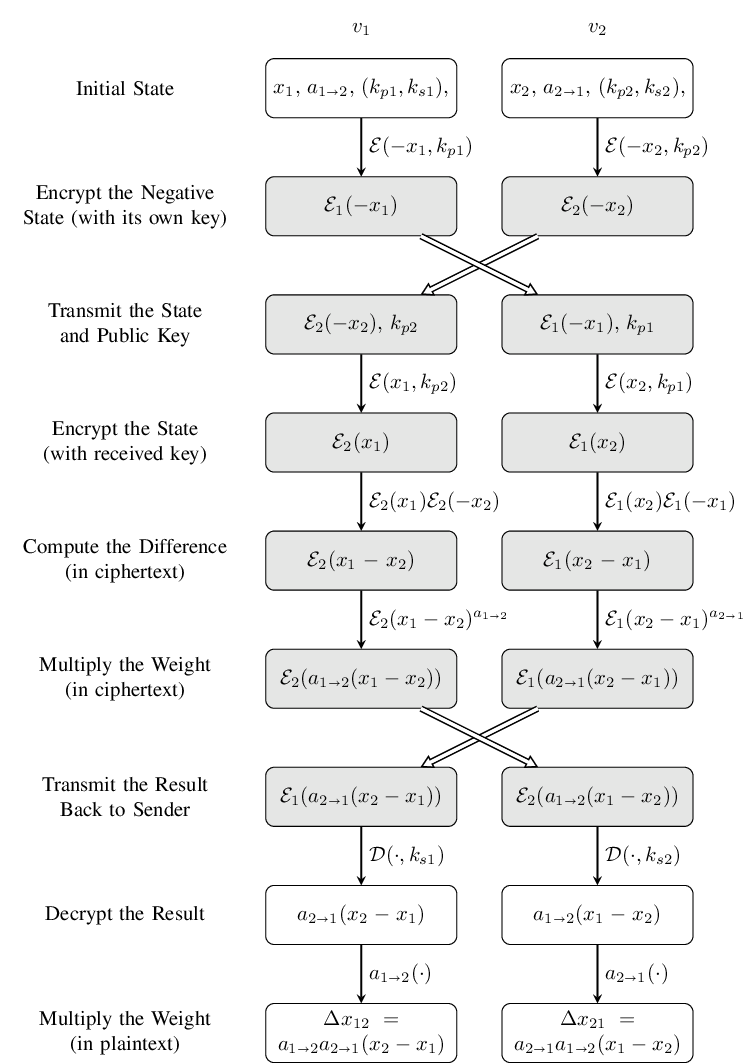}
 \caption{A step-by-step illustration of the confidential interaction protocol. Single arrows indicate the flow of computations; double arrows indicate data exchange via a communication channel. Shaded nodes indicate the computation done in ciphertext. Note that $a_{1 \shortrightarrow 2}$ and $a_{2 \shortrightarrow 1}$ are different from step to step \citep{ruan2019secure}. }
 \label{fig:1}
\end{wrapfigure}

It has been well known that  static average consensus can be achieved if
the network is connected and there exists some $\eta>0$ such that
$\eta\leq a_{ij}[k]<1$ holds for all $k\geq 0$
\citep{nedic2010constrained}.

\paragraph{Privacy in static average consensus} In the static average consensus problem, the sensitive information are the initial values of individual agents. Namely, agent $i$ should avoid its initial value $x_i[0]$  from being inferrable by honest-but-curious adversaries (i.e., other participating agents) and  eavesdroppers (i.e., external observers).

\subsection{Literature review}
In general, existing privacy solutions for the  static average consensus problem  are based on the following mechanisms:

\paragraph{Partially homomorphic encryption based approaches}
Since  commonly used encryption schemes rely on a trusted party to manage encryption and decryption keys, they are not appropriate for fully decentralized multi-agent systems. To the contrary,  homomorphic encryption schemes  allow  computations to be performed on encrypted data without first having to decrypt it, and hence can be implemented in a fully decentralized setting without any trusted party to manage encryption and decryption keys. Homomorphic encryption schemes can be divided into two different categories, fully homomorphic encryption schemes and partially homomorphic encryption schemes.  Although fully homomorphic encryption mechanisms allow  any functions of unbounded depth to be evaluated in the encrypted domain, such approaches are extremely heavy in computation and communication and hence are rarely used in practice. Partially homomorphic encryption schemes can only allow functions of certain types, such as addition or multiplication, to be evaluated in the encrypted domain. However, their communication and communication overheads are manageable in many low-cost computing platforms, making them widely usable in practice. Some of the most popular partially homomorphic encryption schemes include RSA \citep{Rivest1978}, ElGamal \citep{ElGamal1985}, and Paillier \citep{Paillier1999}.

Partially homomorphic encryption was first introduced to the control domain by \cite{Kogiso2015} who first applied partially homomorphic encryption in  a networked control system. Although plenty of results were reported following \cite{Kogiso2015}, there is a major hurdle for applying such approaches in the static average consensus problem, where the interaction weights have to be symmetric in undirected interaction graphs. In fact, in static average consensus, whenever  agent $i$  has access to the value of the interaction term   $a_{ij}[k] (x_j[k]- x_i[k])$ and the interaction weights $a_{ij}[k]$, it can always infer the state value of its neighbor $j$. \cite{ruan2017secure} and \cite{ruan2019secure} first solved the problem by proposing a mechanism to make the interaction weight $a_{ij}[k]$  unknown to both agent $i$ and agent $j$. The idea is to decompose the interaction weight for  any pair of interacting agents into the product  of two positive values which are private to the two agents, respectively. The idea is illustrated in Fig. \ref{fig:1}, where we represent a pair of  interacting agents as agent $v_1$ and agent $v_2$ for the sake of notational simplicity.

 \begin{table}[t]
\caption{Privacy solutions for static average consensus}\label{chap1:tab1}
\begin{center}
\begin{tabular}{ |c|l|m{4cm}|m{5cm}| }
\hline
\multicolumn{2}{|c|}{Privacy mechanisms} & Typical relevant results & Comments \\
\hline
\multirow{2}{11em}{Partially homomorphic encryption} & fully decentralized & \cite{ruan2017secure}, \cite{ruan2019secure}, \cite{hadjicostis2020privacy}, \cite{fang2021secure}, \cite{yin2020accurate}, \cite{yu2021multi}, \cite{gao2021encryption} & Heavy in computation/communication overhead \\
\cline{2-4}
&with a server &\cite{kogiso2015cyber}, \cite{gao2021encryption} & Heavy in computation/communication overhead \\
\hline
\multirow{2}{11em}{Decomposition} & state decomposition & \cite{wang2019privacy}, \cite{wang2021privacy}, \cite{zhang2022privacy}, \cite{zhang2022privacy2}, \cite{chen2023privacy}, \cite{duan2023cooperative}   &   \\
\cline{2-4}
&edge decomposition &\cite{zhang2022privacy}, \cite{xiong2022privacy}   &   \\
\hline
\multirow{2}{11em}{Dynamics based} & directed graph & \cite{gao2018privacy}, \cite{gao2022algorithm}, \cite{gao2022privacy}   & Information theoretic privacy  \\
\cline{2-4}
&undirected graph &  \cite{gupta2019statistical}  &  Information theoretic privacy \\
\hline
\multirow{2}{11em}{Differential privacy} &   
     {decentralized} 
   & \cite{nozari2017differentially}, \cite{he2018privacy}, \cite{gao2018differentially}, \cite{wang_yamin2021differentially},  \cite{fiore2019resilient}, \cite{he2020differential}, \cite{liu2020differentially}, \cite{he2019consensus}, \cite{zhang2022differentially},  \cite{katewa2018privacy}, \cite{zhang2022much}, \cite{chen2023differentially}, \cite{wang2023_yadifferentially}    & Lose accurate convergence  \\
\cline{2-4}
&with a server &  \cite{huang2012differentially}  &  Lose accurate convergence \\
\hline
\multirow{1}{11em}{Observability based} & undirected graph & \cite{manitara2013privacy}, \cite{mo2016privacy}, \cite{kia2015dynamic}, \cite{alaeddini2017adaptive} & Restricted in interaction topology \\
\cline{2-4}
\hline
\end{tabular}
\end{center}
\end{table}

\begin{wrapfigure}{r}{0.6\textwidth}
    \includegraphics[width=0.6\textwidth]{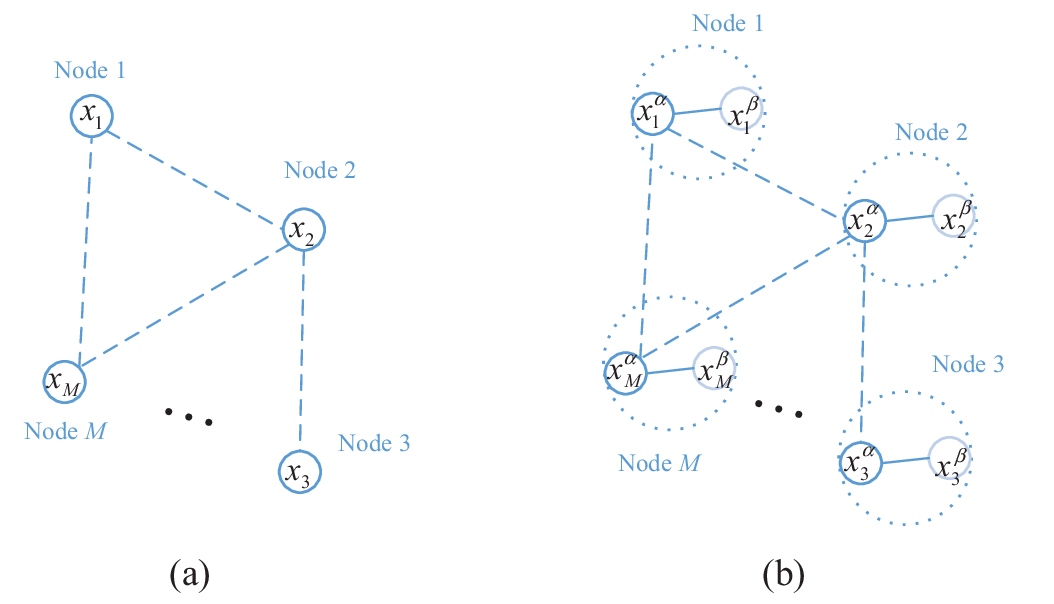}
    \caption{State-decomposition based privacy-preserving  average consensus \citep{wang2019privacy}. (a) Before state decomposition (b) After state decomposition}
    \label{fig:2}
\end{wrapfigure}

\paragraph{Decomposition based approaches} The decomposition based privacy mechanism was first proposed in our work \citep{wang2019privacy}. Its basic idea is to  decompose each agent's state
$x_i$ into two sub-states $x_i^{\alpha}$ and $x_i^{\beta}$, with the
initial values $x_i^{\alpha}[0]$ and $x_i^{\beta}[0]$ randomly
chosen from the set of all real numbers under the constraint
$x_i^{\alpha}[0]+x_i^{\beta}[0]=2x_i[0]$ (see Fig. \ref{fig:2}). The
sub-state $x_i^{\alpha}$ succeeds the role of the original state
$x_i$ in  inter-node interactions and it is in fact the only state
value from node $i$ that can be seen by its neighbors. The other
sub-state $x_i^{\beta}$ also involves in the distributed interaction
by (and only by) interacting with $x_i^{\alpha}$. So the existence
of $x_i^{\beta}$ is invisible to neighboring nodes of node $i$,
although it directly affects the evolution of $x_i^{\alpha}$. Taking
node 1 in Fig. \ref{fig:2}(b) for example, $x_1^{\alpha}$ acts as if
it were $x_1$ in the inter-node interactions while $x_1^{\beta}$ is
invisible to nodes other than node $1$, although it affects the
evolution of $x_1^{\alpha}$.

\paragraph{Dynamics based approaches}There are two types of dynamics based privacy approaches for static average consensus. The first approach employs the robustness of dynamical systems stability to embed uncertainty based privacy without compromising convergence accuracy. For example, we know that for a scalar dynamical system $\dot{x}=ax$ where $x$ is the state and $a$ is a constant,   it is always stable when $a$ is negative, no matter what value $a$ is. Employing this idea, we can introduce uncertainties in the coupling weights judiciously to enable privacy protection without compromising the accuracy of convergence. This idea is first employed in \cite{ruan2017secure,ruan2019secure} with the assistance of encryption and then generalized in \cite{gao2018privacy,gao2022algorithm} without the assistance of encryption. The second dynamics based privacy approach for static average consensus is to add temporally or spatially corrected noises, which dates back at least to \cite{abbe2012privacy}. This approach has been employed for privacy protection in static average consensus in \cite{mo2016privacy},  \cite{manitara2013privacy}, and \cite{gupta2019statistical}, among others.\\

\paragraph{Differential privacy based approaches} Differential privacy is a privacy framework initially proposed for protecting static datasets. Intuitively speaking, differential privacy requires that for a mechanism performed on a dataset, when the dataset is changed in at most one entry, the output distribution of the mechanism is not changed significantly. The most commonly used definition of differential privacy is called $\epsilon$-differential privacy, which is defined as follows \cite{dwork2014algorithmic}:
\begin{Definition 1}
   ($\epsilon$-differential privacy \cite{huang2012differentially}). For a given $\epsilon>0$, a static average consensus  algorithm   is $\epsilon$-differentially private if for any two sets of initial states $\mathcal{P}$ and $\mathcal{P}'$ that differ in at most one agent's initial value (usually called adjacent initial states), any set of observation sequences $\mathcal{O}_s\subseteq\mathbb{O}$ (with $\mathbb{O}$ denoting the set of all possible observation sequences),  we always have
    \vspace{-2mm}
    \begin{equation}
        {\mathbb{P}[\mathcal{R}_{\mathcal{P}} \left(\mathcal{O}_s\right)]}\leq e^\epsilon{\mathbb{P}[\mathcal{R}_{\mathcal{P}'} \left(\mathcal{O}_s\right)]}
    \end{equation}
    where $\mathcal{P}$  denotes the mapping from initial states to observations under a given consensus algorithm and the probability $\mathbb{P}$ is taken over the randomness over iteration processes.
 \end{Definition 1}
 Since  differential privacy is defined under the probabilistic framework, it is usually achieved by injecting additive noises to shared messages. The first differentially private static average consensus approach was proposed in \cite{huang2012differentially} under the assistance of a central server. Fully decentralized solutions for differentially private static average consensus have been proposed in \cite{nozari2017differentially}, \cite{he2018privacy}, and \cite{katewa2018privacy}, among others.

\paragraph{Observation based privacy} This approach  achieves privacy by making a certain state unobservable to some adversarial agents. However, given that the interaction graph has to be connected in static average consensus to ensure that all agents can converge to the same desired value, this approach can only achieve a very limited level of privacy protection.

\section{Privacy protection for dynamic average consensus}

\subsection{Problem formulation}

\paragraph{Dynamic average consensus} We consider a dynamic average consensus problem  among a set of $m$   agents  $[m]=\{1,\,\cdots,m\}$. We index the agents by $1,\,2,\,\cdots,m$. Agent $i$   can access fixed-frequency samples of its own reference signal $r_i\in\mathbb{R}^d$, which could be  varying with time. Every agent $i$ also maintains a state $x_i\in\mathbb{R}^d$. The aim of dynamic average consensus is for all agents to collaboratively track the average reference signal $\bar{r}\triangleq \frac{\sum_{i=1}^{m} {r_i}}{m}$ while every agent can only access discrete-time measurements of its own reference signal and share its state with its immediate neighboring agents.

  We describe the local communication among agents using a weight matrix
$L=\{L_{ij}\}$, where $L_{ij}>0$ if agent $j$ and agent $i$ can directly communicate with each other,
and $L_{ij}=0$ otherwise. For an agent $i\in[m]$,
its  neighbor set
$\mathbb{N}_i$ is defined as the collection of agents $j$ such that $L_{ij}>0$.
We define $L_{ii}\triangleq-\sum_{j\in\mathbb{N}_i}L_{ij}$  for all $i\in [m]$,
where $\mathbb{N}_i$ is the neighbor set of agent $i$.

\paragraph{Privacy in dynamic average consensus}
 In the dynamic average consensus problem, the sensitive information are the reference signals of individual agents.  Namely, we have to make sure that the reference signal $r_i$ of agent $i$ is not inferable by   honest-but-curious adversaries (i.e., other agents participating in the dynamic average consensus
problem) and eavesdroppers (i.e., external observers).
\subsection{Literature review}
Compared with the static average consensus problem, existing results on privacy protection for dynamic average consensus are relatively sparse \citep{zhangk2022privacy}. In fact, given that in many dynamic average consensus problems, the initial state $x_i[0]$ of agent $i$ is usually set as the initial value of the reference signal $r_i[0]$, protecting the reference signal $r_i$ includes protecting initial value as a special case. In fact, protecting the entire signal $r_i$ is equivalent to protecting the values of $r_i$ at infinitely many time instants, which makes privacy protection for dynamic average consensus much more challenging than privacy protection for static average consensus.

It is worth noting that in many applications of dynamic average consensus, such as distributed optimization where $r_i$ is the gradient of agent $i$, many privacy solutions have been proposed. However, since we will specifically discuss privacy protection in distributed optimization in the next section, we do not consider those results in this section.  We want to emphasize the results in \cite{wang2023robust} which proposed a robust dynamic average consensus algorithm that can ensure both differential privacy and accurate convergence:

\noindent\rule{\textwidth}{0.5pt}
\noindent\textbf{Algorithm 1: Robust dynamic average consensus \citep{wang2023robust}}

\noindent\rule{\textwidth}{0.5pt}
\begin{enumerate}[wide, labelwidth=!, labelindent=0pt]
    \item[] Parameters: Weakening factor $\chi^k>0$ and stepsize $\alpha^k>0$.
    \item[] Every agent $i$'s  reference signal is $r_i^k$. Every agent $i$ maintains one state variable  $x_i^k$, which is initialized as $x_i^0=r_i^0$.
    \item[] {\bf for  $k=1,2,\ldots$ do}
    \begin{enumerate}
        \item Every agent $j$ adds persistent DP-noise   $\zeta_j^{k}$ 
        to its state
    $x_j^k$,  and then sends the obscured state $x_j^k+\zeta_j^{k}$ to agent
        $i\in\mathbb{N}_j$.
        \item After receiving  $x_j^k+\zeta_j^k$ from all $j\in\mathbb{N}_i$, agent $i$ updates its state  as follows:
        \begin{equation}\label{eq:update_in_Algorithm1}
        \begin{aligned}
             x_i^{k+1} &=(1-\alpha^k)x_i^k+\chi^k\sum_{j\in \mathbb{N}_i} L_{ij}(x_j^k+\zeta_j^k-x_i^k)+r_i^{k+1}-(1-\alpha^k)r_i^k.
        \end{aligned}
        \end{equation}
    \end{enumerate}
\end{enumerate}
\vspace{-0.1cm} \rule{\textwidth}{0.5pt}

It is worth noting that recently \cite{wang2024robust} extended the result to the constrained consensus case where the state of every agent is constrained in a nonempty, closed, and convex set $X\subset\mathbb{R}^d$ (see details in Algorithm 2). However, it is worth noting that the problem in  \cite{wang2024robust} is not a standard dynamic average consensus problem, since the final convergence point does not necessarily equal to the average reference signal therein.

\noindent\rule{\textwidth}{0.5pt}
\noindent\textbf{Algorithm 2: Differentially-private constrained dynamic consensus \citep{wang2024robust}}

\noindent\rule{\textwidth}{0.5pt}
\begin{enumerate}[wide, labelwidth=!, labelindent=0pt]
    \item[] Parameters: Weakening factor $\chi^k>0$ and stepsize $\gamma^k>0$.
    \item[] Every agent $i$'s  input is $r_i^k$.
    Every agent $i$ maintains one state variable  $x_i^k$, which is initialized randomly in $X$.
    \item[] {\bf for  $k=1,2,\ldots$ do}
    \begin{enumerate}
        \item Every agent $j$ adds persistent DP-noise   $\zeta_j^{k}$ 
        to its state
    $x_j^k$,  and then sends the obscured state $x_j^k+\zeta_j^{k}$ to agent
        $i\in\mathbb{N}_j$.
        \item After receiving  $x_j^k+\zeta_j^k$ from all $j\in\mathbb{N}_i$, agent $i$ updates its state  as follows:
        \begin{equation}\label{eq:update_in_Algorithm1}
        \begin{aligned}
              x_i^{k+1} &
              =\Pi_X\left[ x_i^k+\chi^k{ \textstyle\sum_{j\in \mathbb{N}_i}} w_{ij}(x_j^k+\zeta_j^k-x_i^k)+\gamma^k r_i^k\right].
        \end{aligned}
        \end{equation}
        where $\Pi_X$ denotes  the Euclidean projection to the set $X$.
    \end{enumerate}
\end{enumerate}
\vspace{-0.1cm} \rule{\textwidth}{0.5pt}

\section{Privacy protection for distributed optimization}
\subsection{Problem formulation}
\paragraph{Distributed optimization}We consider  a network of $m$ agents, interacting on a general directed graph. We describe a directed graph using an ordered pair $\mathcal{G}=([m],\mathcal{E})$, where
$[m]=\{1,2,\ldots,m\}$ is the set of nodes (agents) and $\mathcal{E}\subseteq [m]\times [m]$  is the edge set of ordered node pairs describing the interaction among agents.
For a nonnegative weighting matrix $L=\{L_{ij}\}\in\mathbb{R}^{m\times m}$, we define the induced directed graph as $\mathcal{G}_L=([m],\mathcal{E}_L)$, where
the directed edge $(i,j)$ from agent $j$ to agent $i$ exists,
 i.e., $(i,j)\in \mathcal{E}_L$ if and only if $L_{ij}>0$.
For an agent $i\in[m]$,
its in-neighbor set
$\mathbb{N}^{\rm in}_i$ is defined as the collection of agents $j$ such that $L_{ij}>0$; similarly,
the out-neighbor set $\mathbb{N}^{\rm out}_i$ of agent $i$ is the collection of agents $j$ such that $L_{ji}>0$.

The distributed optimization problem  can be
reformulated as follows:
\begin{equation}\label{eq:optimization_formulation1}
\min\limits_{\theta\in\mathbb{R}^d} F(\theta)\triangleq
\frac{1}{m}\sum_{i=1}^m f_i(\theta)
\end{equation}
where $m$ is the number of agents, $\theta\in\mathbb{R}^d$ is
 a decision variable common to all agents, while
$f_i:\mathbb{R}^d\rightarrow\mathbb{R}$ is a local objective
function private to agent $i$.

It is worth noting that when the local objective function $f_i$ is set as $f_i=\|\theta-\theta_i[0]\|^2$, then the above distributed optimization problem reduces to the static average consensus problem \citep{zhang2018enabling}.

\paragraph{Privacy in distributed optimization} In most applications of distributed optimization, the sensitive information are contained in the objective function or gradient of participating agents. For example,
  in sensor network based target localization, the positions of sensors should  be kept private in sensitive (hostile) environments  ~\citep{zhang2019admm,huang2015differentially}. In existing distributed optimization based localization   algorithms, the position  of a sensor is a parameter of its objective function, and as shown in ~\cite{zhang2019admm,huang2015differentially,burbano2019inferring}, it is easily inferable by an adversary  using  information shared in these distributed algorithms. The privacy problem is more acute in distributed machine learning where involved training data may contain sensitive information such as medical or salary information (note that in machine learning, together with the model, training data  determines the objective function). In fact, as shown in our recent results \citep{wang2022quantization,wang2022tailoring,wang2022decentralized}, in the absence of  a   privacy mechanism, an adversary can use information shared  in distributed optimization to precisely recover the raw data used for training.

\subsection{Literature review}

In Table \ref{chap1:optimization}, we summarize typical existing results on privacy protection for distributed optimization. It is worth noting that since we focus on decentralized optimization, many other results based on cloud/server (see, e.g.,  \cite{xiong2020privacy})  are not included.

\begin{table}[t]
\caption{Privacy solutions for distributed optimization}\label{chap1:optimization}
\begin{center}
\begin{tabular}{ |c|l|m{4cm}|m{5cm}| }
\hline
\multicolumn{2}{|c|}{Privacy mechanisms} & Typical relevant results & Comments \\
\hline
\multirow{2}{11em}{\\Partially homomorphic encryption} & fully decentralized & \cite{zhang2018admm}, \cite{zhang2018enabling}  & Heavy in computation/communication overhead \\
\cline{2-4}
&with a server &\cite{lu2018privacy}, \cite{alexandru2020cloud} & Heavy in computation/communication overhead \\
\hline
\multirow{2}{11em}{Decomposition} & state decomposition & \cite{zhang2018privacy}, \cite{chen2023differentially}, \cite{sun2023privacy}   &   \\
\hline
\multirow{3}{11em}{Dynamics based} &  coupling weight based & \cite{zhang2018improving}, \cite{gao2023dynamics},     &   \\
\cline{2-4}
&stepsize based &  \cite{wang2022decentralized}, \cite{wang2023decentralized}  & \cite{wang2022decentralized} achieved information theoretic privacy \\
\cline{2-4}
&quantization based &  \cite{wang2022quantization}  &  Achieved differential privacy \\
\hline
\multirow{2}{11em}{\\Differential privacy} & decentralized & \cite{Huang15}, \cite{zhang2016dynamic}, \cite{ding2021differentially},\cite{wang2022tailoring}, \cite{xuan2023gradient}, \cite{wang2023decentralized_auto}, \cite{nozari2016differentially}, \cite{mao2023differentially}, \cite{wu2022differentially}, \cite{zhao2022differential}    & \cite{wang2022tailoring} maintains accurate convergence while ensuring differential privacy \\
\cline{2-4}
&with a server &  \cite{han2016differentially}, \cite{hale2017cloud}  &  Lose accurate convergence \\
\hline
\end{tabular}
\end{center}
\end{table}

\subsection{Typical algorithms}
\noindent\rule{\textwidth}{0.5pt}
\noindent\textbf{Algorithm 3: Differential-privacy-oriented  distributed optimization}

\noindent\rule{\textwidth}{0.5pt}
\begin{enumerate}[wide, labelwidth=!, labelindent=0pt]
    \item[] Parameters: Stepsize $\lambda^k$ and
    weakening factor $\gamma^k$.
    \item[] Every agent $i$ maintains one state     $x_i^k$, which is initialized with a random vector in $\mathbb{R}^d$.
    \item[] {\bf for  $k=1,2,\ldots$ do}
    \begin{enumerate}
        \item Every agent $j$ adds persistent DP-noise   $\zeta_j^{k}$ 
        to its state
    $x_j^k$,  and then sends the obscured state $x_j^k+\zeta_j^{k}$ to agent
        $i\in\mathbb{N}_j^{\rm out}$.
        \item After receiving  $x_j^k+\zeta_j^k$ from all $j\in\mathbb{N}_i^{\rm in}$, agent $i$ updates its state  as follows:
        \begin{equation}\label{eq:diminishing_update}
        \hspace{-0.5cm}
        \begin{aligned}
             x_i^{k+1}\hspace{-0.1cm}&=\hspace{-0.1cm}x_i^k+\sum_{j\in \mathbb{N}_i^{\rm
            in}}\gamma^k L_{ij}(x_j^k+\zeta_j^k-x_i^k)-\lambda^k\nabla f_i(x_i^k)
        \end{aligned}
        \end{equation}
                \item {\bf end}
    \end{enumerate}
\end{enumerate}
\vspace{-0.1cm} \rule{\textwidth}{0.5pt}

 The sequence $\{\gamma^k\}$ diminishes with time and is used to suppress the influence of persistent differential-privacy noise $\zeta_j^k$ on the convergence point of the iterates.
The stepsize sequence $\{\lambda^k\}$ and attenuation sequence $\{\gamma^k\}$
have to be designed appropriately to guarantee the almost sure convergence of all $\{x_i^k\}$   to a common optimal solution $\theta^{\ast}$.
The persistent differential-privacy  noise processes $\{\zeta_i^k\}, i\in[m]$ have zero-mean its variance is allowed to increase with time. In fact, allowing the variance to increase with time is key for our approach to enabling rigorous differential privacy while maintaining accurate convergence, even in the infinite time horizon. It is worth noting that an increasing noise variance   will make the relative level between noise $\zeta_i^k$ and signal $x_i^k$ increase  with time. However, since the increase  in noise variance can be outweighed by the decrease of $\gamma^k$, the actual noise fed into the algorithm, i.e., $\gamma^k\zeta_j^k$, still decays with time, which makes it possible for  Algorithm 3 to ensure almost sure convergence to an optimal solution.

\section{Privacy protection for other algorithms in multi-agent systems}
We considered privacy protection in  static average consensus, dynamic average consensus, and distributed optimization, which  are the three most important primitives for coordination in multi-agent systems. In fact, the problem of privacy protection has also been addressed in many other  algorithms for multi-agent systems. For example, distributed Nash equilibrium seeking is receiving increased  traction in recent years due to its ability to capture the noncooperative relationship among agents in many multi-agent systems. To enable privacy protection in distributed Nash equilibrium seeking, plenty of efforts have been reported (see, e.g., \cite{ye2021differentially,wang2022differentially_jimin}). Two specific results worth mentioning are  our recent results in \cite{wang2022ensuring} and \cite{wang2024differentially} which enable  differential privacy and accurate convergence simultaneously in aggregative games and general games, respectively. In addition, bipartite consensus is   an algorithm for multi-agent systems which can model the dynamics in social networks. Recently, \cite{zuo2022differential} and \cite{wang2024differentially_jimin} studied differential privacy for bipartite consensus. Furthermore, broadly speaking, networked control systems \citep{yong2008fault} and oscillator networks \citep{wang2011influences} can also be viewed as multi-agent systems (with heterogeneous agents and continuous-time interactions, respectively). Their privacy protection problem is also gaining increased attention recently \citep{cortes2016differential,gupta2018model,sultangazin2018protecting,darup2021encrypted,rezazadeh2018privacy}.

\section{Typical Applications}

\subsection{Application in robot networks}
\begin{wrapfigure}{r}{0.55\textwidth}
	\begin{center}
		\vspace{-0.1cm}\includegraphics[width=0.45\textwidth]{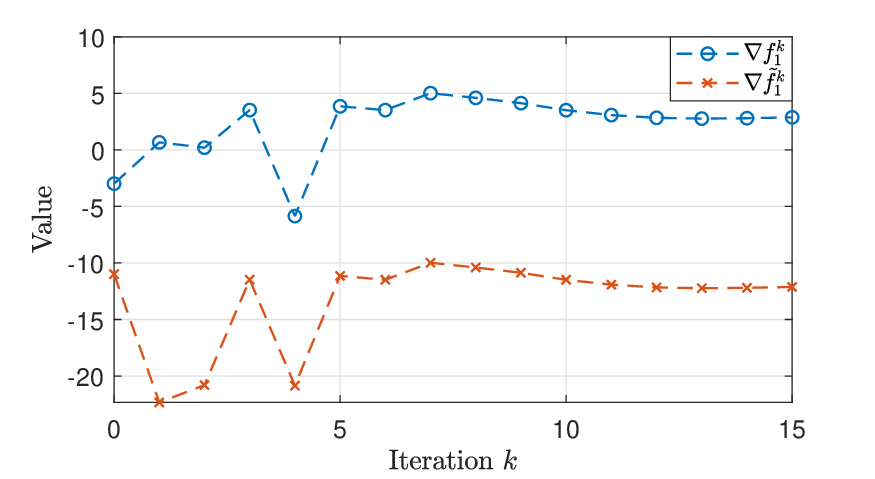}
	\end{center}
	\vspace{-0.2cm}\caption{The two different gradient functions of node $1$ that lead to identical observations \citep{gao2022dynamics}.}
	\label{fig_gradient_comparison}
\end{wrapfigure}

We consider the distributed rendezvous problem where a group of robots want to agree on the nearest meeting point without revealing each other's trajectories \citep{huang2015differentially} (note that the position information of a robot is embedded in its local gradient function). Mathematically, this can be modeled as the problem $\min_{x\in \mathbb{R}^d} \, \ \sum_{i=1}^{m} f_i(x) = \sum_{i=1}^{m} \frac{1}{2} \| x- {p}_i\|^2$, where $ {p}_i$ represents the initial position of node $i$. For the simplicity of exposition, we consider the $d=1$ case but similar results can be obtained when $d\neq 1$. We consider a circle graph where an agent can only communicate with its two immediate neighbors.  We use the privacy approach in \cite{gao2022dynamics} which employs uncertainties in inter-agent coupling to make one agent's gradient indistinguishable by adversaries from observations (shared information). Fig. \ref{fig_gradient_comparison} shows the two different gradients of agent 1 that can lead to the same observations, which clearly makes agent 1's gradients indistinguishable by adversaries.

\subsection{Application in machine learning}

\begin{figure}[!htb]
    \centering
    \begin{minipage}{.5\textwidth}
        \centering
        \includegraphics[width=0.5 \textwidth]{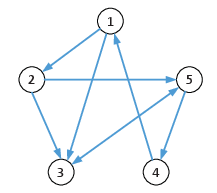}
        \caption{The interaction graph.}
        \label{fig:topology}
    \end{minipage}%
    \begin{minipage}{0.5\textwidth}
        \centering
        \includegraphics[width=0.8\textwidth]{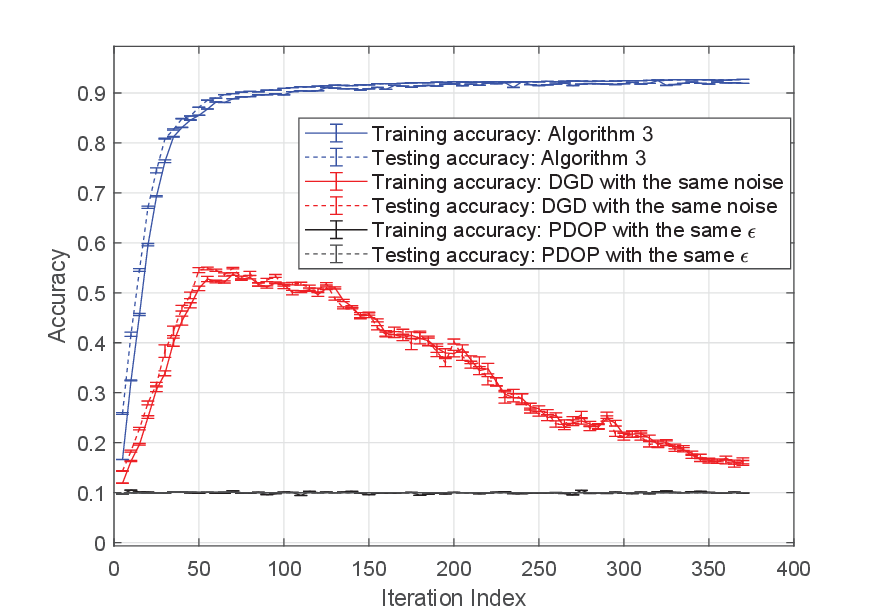}
        \caption{Comparison of Algorithm 1 in \cite{wang2022tailoring} with the distributed gradient descent algorithm (DGD) in \cite{nedic2009distributed} (under the same noise) and the differential-privacy approach for decentralized optimization PDOP in \cite{huang2015differentially} (under the same privacy budget) using the MNIST image classification problem}
        \label{fig:mnist}
    \end{minipage}
\end{figure}

We consider the decentralized training of a convolutional neural network
(CNN).
More specially, we   consider five agents which collaboratively
train a CNN using the MNIST dataset \citep{MNIST} under the topology
 in Fig. \ref{fig:topology}. The MNIST  data set is a large
benchmark database of handwritten digits widely used for training
and testing in the field of machine learning \citep{deng2012mnist}.
Each agent has a local copy of the CNN. The CNN has 2 convolutional
layers with 32 filters with each followed by a max pooling layer,
and then two more convolutional layers with 64 filters each followed
by another max pooling layer and a dense layer with 512 units. Each
agent has access to a portion of the MNIST dataset, which was
further divided into two subsets for training and validation,
respectively. We use the differentially private Algorithm 3  \citep{wang2022tailoring} to enable privacy, where  the stepsize was set as  $\lambda^k= \frac{1}{1+0.01k}$ and the weakening factor was set as $\gamma^k$ as $\frac{1}{1+0.01k^{0.9}}$. The Laplace noise parameter was set to $\nu^k=1+0.01k^{0.3}$ to enable $\epsilon$-differential privacy.
The evolution of the training and testing accuracies averaged
over 50 runs are illustrated by the solid and dashed blue curves
in Fig. \ref{fig:mnist}. To compare the convergence performance
of this algorithm with the conventional distributed
gradient descent algorithm under differential privacy noise, we also show the results of using the distributed
 gradient descent (DGD)  algorithm in \cite{nedic2009distributed} to train
the same CNN {using stepsize $\frac{1}{1+0.01k}$} under the same Laplace noise. The results are illustrated by the solid and dotted red curves in Fig. \ref{fig:mnist}. It can be seen that  Algorithm  3 has  much better robustness to  differential privacy noise.  Moreover, to compare with the differential privacy approach for distributed optimization (PDOP) in \cite{Huang15}, we also plot the results under PDOP in \cite{Huang15}    under the same privacy budget $\epsilon$. PDOP uses geometrically decaying stepsizes and noises to ensure a finite privacy budget. However,   such fast-decaying stepizes  turned out to be unable to train the complex CNN model (see training and testing accuracies in solid and dashed black curves in Fig. \ref{fig:mnist}, respectively under   $\lambda^k=0.95^k$ and $\nu^k=0.98^k$).

\section{Conclusions}

We have discussed several typical approaches for privacy protection in multi-agent systems. In fact, all of the discussed results with superior performances are based on some kind of co-design of the privacy mechanism and coordination algorithms.  Although different approaches have their respect advantages and disadvantages, and new privacy results have been continuously emerging from the control domain,  we believe that only by cross fertilizing  privacy results in computer science and control can we ensure effective privacy protection in multi-agent systems while retaining real-time and accuracy guarantees of coordination algorithms, which are essential for promoting multi-agent system applications in  practical domains such as power systems and intelligent transportation.

\bibliographystyle{Harvard}
\bibliography{reference}

\end{document}